\documentclass[11pt]{article}
\usepackage{graphicx}
\usepackage{epstopdf}
\usepackage[square,sort,comma,numbers]{natbib}
\usepackage{setspace}
\usepackage{color}

\title{Quasi-periodic acceleration of electrons by a plasmoid-driven shock in the solar atmosphere}

\author{Eoin P. Carley$^{1}$, David M. Long$^2$, Jason P. Byrne$^3$, Pietro Zucca$^1$,\\ D. Shaun Bloomfield$^1$, Joseph McCauley$^1$, \& Peter T. Gallagher$^{1,*}$}
\smallskip

\begin{document}

\maketitle

\begin{enumerate}
 	\item {\small Astrophysics Research Group, School of Physics, Trinity College Dublin, Dublin 2, Ireland.} \and
 	\item {\small Mullard Space Science Laboratory, University College London, Holmbury St. Mary, Dorking, Surrey, RH5 6NT, UK.}\and
	\item {\small Institute for Astronomy, University of Hawai'i, 2680 Woodlawn Drive, Honolulu, HI 96822, USA.} \\
	$^{*}$\emph{{\small peter.gallagher@tcd.ie}}
\end{enumerate}

\begin{abstract}
Cosmic rays and solar energetic particles may be accelerated to relativistic energies by shock waves in astrophysical plasmas.
%
On the Sun, shocks and particle acceleration are often associated with the eruption of magnetized plasmoids, called coronal mass ejections (CMEs).
However, the physical relationship between CMEs and shock particle acceleration is not well understood.
Here, we use extreme ultraviolet, radio and white-light imaging of a solar eruptive event on 22 September 2011 to show that a CME-induced shock (Alfv\'{e}n Mach number $2.4^{+0.7}_{-0.8}$) was coincident with a coronal wave and an intense metric radio burst generated by intermittent acceleration of electrons to kinetic energies of 2--46\,keV (0.1--0.4\,c). 
Our observations show that plasmoid-driven quasi-perpendicular shocks are capable of producing quasi-periodic acceleration of electrons, an effect consistent with a turbulent or rippled plasma shock surface.
\end{abstract}


Coronal mass ejections (CMEs) are spectacular eruptions of magnetized plasma from the low solar atmosphere into interplanetary space \citep{byrne2010, roussev2012}. With kinetic energies of $\sim$$10^{25}$\,J \citep{vourlidas2010}, they are the most energetic explosive events in the solar system and are often associated with plasma shocks and the acceleration of particles to relativistic speeds \citep{klassen2002, grechnev2011}. However, the underlying mechanism relating CMEs, shocks, and particle acceleration is still a subject of intense debate \citep{vrsnak2008}. By clarifying the inherent characteristics of these phenomena we learn not only about the nature of explosive plasma events but also about how they drive shocks and accelerate particles to high energies. Such processes are ubiquitous in the universe, playing a role in the acceleration of cosmic rays in supernovae and active galactic nuclei shocks \citep{drury2012}.

CME-associated shocks are often observed over a variety of spectral bands. At radio frequencies, high intensity ($\sim$$10^8$\,Jy) emissions, known as type II and type III bursts, are associated with coronal shocks and accelerated particles in the solar corona \citep{wild1950, mann1996}. Fine structure in these radio bursts can often reveal a \textquoteleft bursty' nature to the shock particle acceleration \citep{mann2005}, which can reveal details of the internal shock structure \citep{zlobec1993, guo2010}.
At extreme ultraviolet (EUV) wavelengths, the shock or pressure pulse response of the corona to an eruption may be imaged as a bright pulse propagating across the entire solar disk at typical velocities of 200--400\,km\,s$^{-1}$ \citep{gallagher2011}. These \textquoteleft coronal bright fronts' (CBFs) are a regular feature of solar eruptive events and often display wave-like properties such as reflection \citep{gopal2009}, refraction \citep{wang2000} and pulse broadening \citep{long2011}. %
Like CMEs, CBFs are often accompanied by type II and type III radio bursts, with EUV and radio images revealing a spatial link between the phenomena that is suggestive of a common origin \citep{maia2004, kozarev2011, vrsnak2005}.

It has been proposed that the common origin for these myriad phenomena may be a CME-driven shock \citep{grechnev2011, warmuth2004b}.  In this scenario, the CME eruption drives a pressure pulse, observable in the low corona as a propagating wave-like CBF. Higher in the corona this same pulse forms a shock, accelerating particles and producing type II and III emission. However, much debate surrounds the suggestion that (i) the CBF is a plasma pressure wave driven by a CME, and (ii) the radio bursts, generated by accelerated particles, result from this same wave/shock system. The contention has arisen from attempts to explain non-wave kinematics of CBFs \citep{zhukov2009}. Pseudo-wave theories are employed to describe this behavior, where the erupting CME produces a large-scale restructuring of the coronal magnetic field, which results in a propagating bright pulse (via Joule plasma heating) that is not actually a driven wave \citep{delannee2008}. In this scenario, any relationship with shock observables is indirect. Further confusion is added by the possibility that high energy particles in association with the eruption may be a consequence of magnetic reconnection in the flaring active region, and not the result of a shock \citep{kahler2007}.

Collectively, CMEs, CBFs and radio bursts provide direct measures of both shock kinematics and the characteristics of the accompanying accelerated particles. However, a common theory explaining these phenomena has yet to be verified.
This lack of clarity can be ascribed to an EUV imaging cadence that was unable to match the fast time sampling of radio imaging and spectroscopy. Now, using the high image cadence of the Solar Dynamics Observatory \citep[SDO;][]{presnell2012}, combined with fast time sampling radio images and spectra, we can reveal previously unseen characteristics of the relationship between these phenomena, proving that a CME-driven shock is the feature unifying these observations and that this shock is responsible for bursty electron acceleration. This greatly advances our understanding of the close relationship between solar eruptions, plasma shocks and their resulting EUV, radio and particle acceleration signatures.


\section*{Coronal Bright Front and Radio Source}
\begin{figure*}
\begin{center}
\includegraphics[scale=0.6, angle=0, trim=3cm 2cm 1cm 4cm]{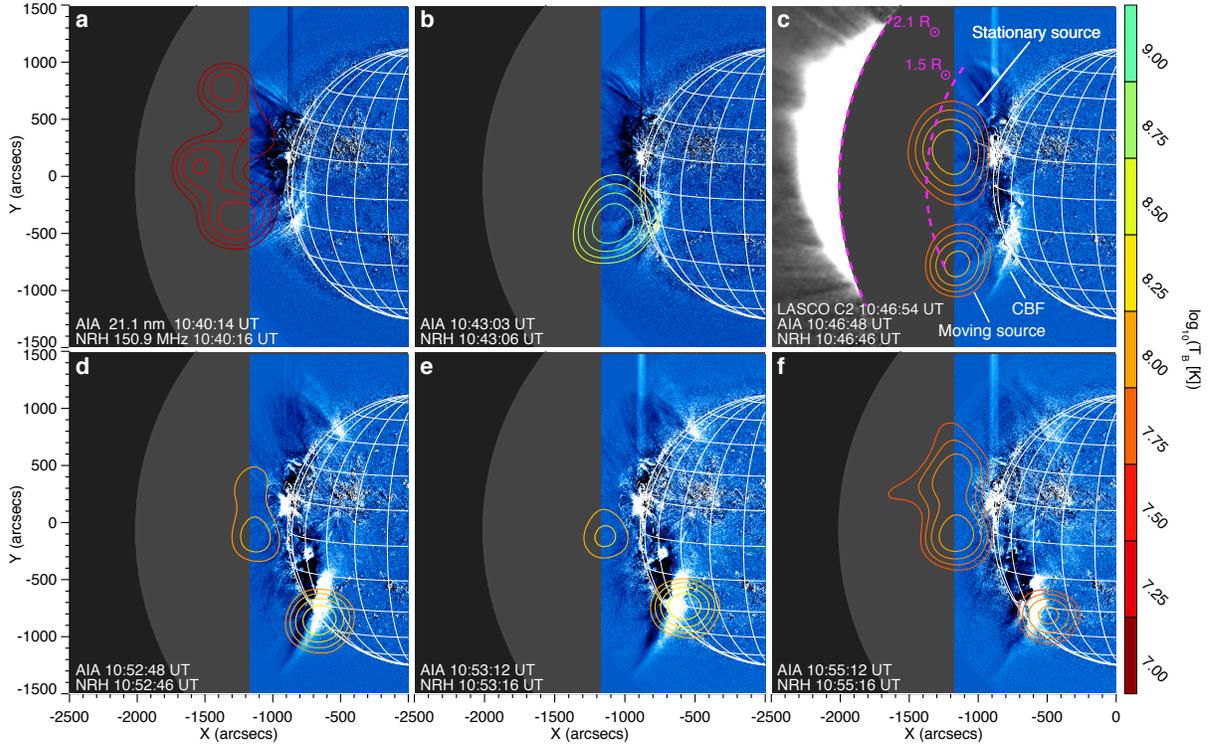}
\caption {AIA \,21.1\,nm images over-plotted with NRH 150.9\,MHz contours. {\bf a-f} show that the 150\,MHz source follows closely the coronal bright front (CBF) as it propagates around the east limb, indicating they belong to a common structure. The intensity of the radio source is indicated by the colour bar on the right, showing the brightness temperature of the source ranges between $\sim$10$^7$--$10^9$\,K. Such high intensities are indicative of coherent plasma emission produced via high energy electron beams. {\bf c} reveals the role of the CME in the event, as observed by the LASCO C2 coronagraph. The combination of the white-light coronagraph (C2) and the EUV images (AIA) reveal the full spatial extent of the CME bubble i.e., the frontal structure in white-light has clear extensions back toward the solar surface, imaged at EUV. The location of the radio source and CBF show they clearly have a relationship with the southward CME flank. The dashed pink lines indicate the predicted height range of the radio emissions observed in the NDA dynamic spectrum (Fig.~\ref{fig:dyn_spec}b). A movie of this figure is available in Supplementary Movie 1.}
\label{fig:figure_aia_nrh_c2}
\end{center}
\end{figure*}

On 22 September 2011 at 10:29\,UT, an X-ray flare (GOES class X1.4) began in an active region located on the east limb of the Sun (NOAA active region 11302; N13E78). Approximately 11 minutes after the flare start time, a bright wave-like front (CBF) was observed propagating away from the southern edge of the active region in SDO/Atmospheric Imaging Assembly \citep[AIA;][]{lemen2012} 21.1\,nm passband images. The CBF then propagated along the east limb from $\sim$15$^{\circ}\,$\,south to $\sim$50$^{\circ}\,$\,south of the equator (Fig.~\ref{fig:figure_aia_nrh_c2}). During the same period of the CBF propagation, a bright 150.9\,MHz source formed above the CBF, imaged using the Nan\c{c}ay Radioheliograph \citep[NRH;][]{kerdraon1997} (contours Fig.~\ref{fig:figure_aia_nrh_c2}). In each image the contours range from $T_{peak}$ to $0.95T_{peak}$, where $T_{peak}$ is the peak brightness temperature at the time of the image; the intensity of the contours is indicated by the colour bar on the right. Initially, both the erupting structure seen in the AIA image and the radio source had the same spatial extent over latitude (Fig.~\ref{fig:figure_aia_nrh_c2}a), showing they belong to a common structure. After this, the most southern part of the radio source reached an extremely high brightness temperature ($\sim$10$^9$\,K) and closely followed the propagation of the CBF southward until it eventually diminished into the thermal background at 10:56\,UT. Another emission source at 150.9\,MHz was also observed at $\sim$$0^{\circ}$ latitude on the east limb at a height of 1.1--1.3$\,R_{\odot}$ (Fig.~\ref{fig:figure_aia_nrh_c2}c,d,e,f); while this source was clearly associated with the eruptive active region, any link between it and the CBF is secondary, as it shows no temporal relationship with the start and stop time of the bright front. Similar radio source motion is observed at 173, 228 and 270\,MHz, however any co-propagation of these radio sources with the CBF is of much shorter duration. A movie showing the co-propagation of the CBF and 150\,MHz radio source can be found in Supplementary Movie 1. For a movie and discussion of the multi-thermal nature of the CBF see Supplementary Movie 2.

\begin{figure*}
\begin{center}
\includegraphics[scale=0.6, angle=0, trim=1cm 1cm 0cm 2cm]{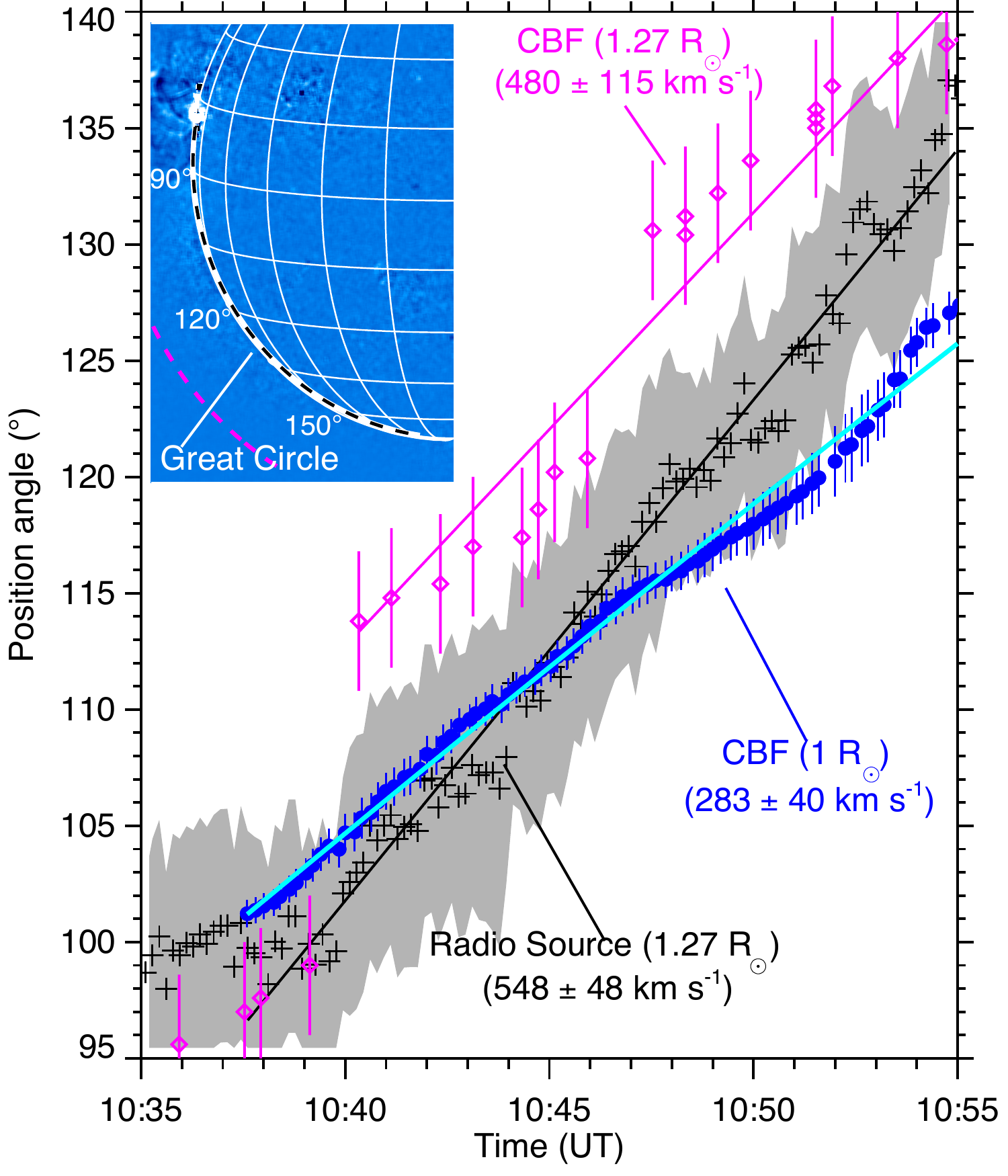}
\caption {Position angle (degrees anticlockwise from solar north) versus time for the 150\,MHz source, shown in plus signs, and coronal bright front (CBF) at $1\,R_{\odot}$ (circles) and $1.27\,R_{\odot}$ (diamonds). The great circle along which the CBF was tracked at $1\,R_{\odot}$ is indicated by the dashed white line in the inset; the dashed pink line marks a height of $1.27\,R_{\odot}$. Both radio burst and CBF have a consistent propagation in the same direction and have similar speeds at a height of $1.27\,R_{\odot}$, implying they belong to a common propagating coronal structure. The uncertainty on radio source position angle is taken to be from 1$\sigma$ uncertainties of the source width ($\sim$7$^{\circ}$) plus the fluctuation of source position due to coronal and ionospheric scattering effects ($3^{\circ}$ at frequencies up to 160\,MHz \citep{stewart1982}). The CBF position uncertainty is from Gaussian centroid uncertainty from a tracking and fitting algorithm of the CBF pulse \citep{long2011a}.}
\label{fig:angle_time}
\end{center}
\end{figure*}

To compare the motion of the CBF and radio source, the position angle (PA) trajectories were analyzed (Fig.~\ref{fig:angle_time}). Both the CBF and radio source clearly show common kinematics, with the two features having a consistent progression southward around the east limb. The solid lines show a fit of $\theta(t) = \theta_0 + \omega t$ to the data, where $\theta_0$ is the starting PA, $\omega$ is the angular velocity, and $t$ is time. The slope of each line gives $\omega$, from which the velocity of the source may be obtained using $v=r\omega$, where $r$ is the distance of the source from Sun center. 
For the CBF, an angular velocity of $4.1\pm0.4\times10^{-4}$\,rad\,s$^{-1}$ was obtained, resulting in a velocity of $283\pm40$\,km\,s$^{-1}$ at $1\,R_{\odot}$.
At $1.27\,R_{\odot}$, the CBF was found to have an angular velocity of $5.4\pm1.3\times10^{-1}$\,rad\,s$^{-1}$, resulting in $v=480\pm115$\,km\,s$^{-1}$.
The 150\,MHz source had an angular velocity of $6.2\pm0.1\times10^{-4}\,\mathrm{rad\,s^{-1}} $; the value for $r$ of this source was estimated by directly converting frequency $f_p$ to electron density $n_e$, using the methods described in Supplementary Material. This gave a height of $1.27^{+0.06}_{-0.09}\,R_{\odot}$ for the radio source, resulting in a velocity of $548^{+34}_{-48}$\,km\,s$^{-1}$. This velocity is larger than the CBF velocity at $1\,R_{\odot}$, but is comparable to the CBF velocity at $1.27\,R_{\odot}$ (Fig.~\ref{fig:angle_time}). 
The similar speeds and trajectories of the CBF and radio source suggest that they belong to a common propagating structure in the corona.
Finally, the radio source motion speed was used to estimate its Alfv\'{e}n Mach number $M_A = v/v_A$, where $v_A$ is the Alfv\'{e}n speed determined from magnetic field and density measurements at the radio source height (see Supplementary Material). With $v_A=225^{+85}_{-35}$\,km\,s$^{-1}$, the Alfv\'{e}n Mach number of the radio source is $2.4^{+0.7}_{-0.8}$, showing that the source travelled in excess of the local wave speed in the corona. Finally, we note that a potential field source surface extrapolation (PFSS) reveals the presence of open and radial field in the south east quadrant of the corona (see Supplementary Figure 2), revealing that the propagation was transverse to the magnetic field. This is an important aspect of quasi-perpendicular shock orientation that we discuss in the last section.


\subsection*{Radio Dynamic Spectra}

The 150\,MHz source observed by NRH had a brightness temperature of $\sim$$10^9$\,K, indicating coherent plasma emission. Such emission is generated via plasma oscillations that are due to instabilities in the presence of high velocity electron beams \citep{dulk1985}. The presence of electron beams was independently verified and revealed in detail using radio dynamic spectra. At $\sim$10:40\,UT the fundamental and harmonic bands of a type II burst shock signature was observed at 45 and 90\,MHz (Fig.~\ref{fig:dyn_spec}b), respectively, using the Nan\c{c}ay Decametric Array \citep[NDA;][]{boischot1980}. Type III bursts begin at the same time as the type II (Fig.~\ref{fig:dyn_spec}a,b), observed using NASA's STEREO-B/WAVES instrument \citep{bougeret2008}. 
The frequency drift of these type III radio bursts provide a measure of velocity of the electrons causing the radio emission. Converting frequency-time measurements to height-time via a coronal density model (see Supplementary Material),  the beam speed was estimated to be $\leq$0.4\,c (46\,keV), where c is the speed of light. This is a clear signature of relativistic electron acceleration in association with the shock. These particles were eventually detected in-situ by the STEREO-B spacecraft (see Supplementary Figure 3).

More striking evidence for shock-accelerated electrons is in the form of `herringbone' emission, observed using the eCallisto spectrometers \citep{benz2009} at the Rosse Solar Terrestrial Observatory \citep[RSTO;][]{zucca2012} (Fig.~\ref{fig:dyn_spec}c). The herringbones result from individual beams of shock-accelerated electrons \citep{mann2005}, traveling towards and away from the Sun  i.e., to higher and lower frequencies. Similar features occur between 100--200\,MHz (Fig.~\ref{fig:dyn_spec}b), showing the same characteristics as herringbones (a bursty nature and decreasing intensity with respect to time). In a similar manner to the type III bursts, the beam velocity of herringbone electrons was estimated to be 0.15\,c, again showing the presence of near-relativistic electron acceleration in association with the presence of a shock. While their structure in frequency reveal how fast these beams travel, their behavior in time can reveal detailed temporal characteristics of the shock acceleration process.
A wavelet analysis using a derivative of a Gaussian wavelet performed on a time series at 54\,MHz reveals periodicity at 2--11~seconds (see Supplementary Figure 1). Previous authors have attributed this bursty nature to rippling and inhomogeneity along the shock front, possibly revealing some level of instability or shock turbulence in the acceleration region \citep{burgess2006, guo2010}; we discuss this in the last section. We note that the features at 100--200\,MHz appear to be the extension of the herringbones into higher frequencies. These features in particular show good temporal correspondence with the radio source i.e., they have a start-stop time comparable to the radio source. This is particularly apparent for the group of bursts at 10:52--10:56\,UT (Supplementary Movie 1).

The radio emission in the dynamic spectra have all the hallmarks of shock generation with particle acceleration closely tied to the process. 
The association of shock radio activity with the imaged 150\,MHz source suggests that the two observables have a common origin in a plasma shock. Overall, the position of the radio source at the southern flank of the CME, the transverse motion of the source (propagation parallel to the surface) and the zero frequency-drift of the herringbones is suggestive of a shock driven parallel to the surface by the flank expansion, similar to the assertion by \cite{stewart1980} and \cite{schmidt2012}. Indeed the association of the CBF with this radio activity is corroborative evidence of the wave/shock system at the flank. Further evidence of a shock having occurred in the corona was obtained through white-light observations, allowing us to study the position of this shock relative to the CME.


\subsection*{White-light CME and Shock}
A CME associated with this event was observed by the Large Angle Spectroscopic Coronagraph \citep[LASCO;][]{bru95}, first appearing at 10:46\,UT, with an apex heliocentric distance of $\sim$2.6\,$R_{\odot}$ (Fig.~\ref{fig:figure_aia_nrh_c2}c). The next available image shows the bright CME front with a fainter, secondary front at the southern flank (Fig.~\ref{fig:wl_shock}b). This `two-front' morphology is a common occurrence in white-light CME structure and constitutes a reliable signature of a CME front associated with a stand-off shock \citep{vourlidas2012}. In order to distinguish between the CME front and shock front, we performed a 3D reconstruction of the CME using the elliptical tie-pointing method described in \cite{byrne2010} (Fig.~\ref{fig:wl_shock}d). 
This reconstruction reveals that the bright front outlined in the C2 coronagraph (ellipse in Fig.~\ref{fig:wl_shock}b) corresponds to the faint front outlined as a halo in STEREO-B COR1 (ellipse in Fig.~\ref{fig:wl_shock}c). 
Furthermore, the observations reveal that the secondary and extremely faint front at the southern edge of the CME (as imaged in LASCO/C2,~Fig.~\ref{fig:wl_shock}b) cannot be considered as part of the CME structure, but is actually an associated shock front. We note that white-light shocks have been reported in the past, occurring both in the low corona as well as out to $\sim$0.5\,A.U. \citep{vourlidas2012, maloney2011}. Here, we have employed a 3D reconstruction from multi-viewpoint observations to qualitatively confirm the presence of a shock at the southern flank of the CME, in the same region as the CBF and radio burst.

\clearpage


\section*{Plasma Shocks and Bursty Particle Acceleration}

There has been much debate surrounding the assertion that CBFs are a wave phenomenon \citep{gallagher2011}, with numerous authors suggesting a pseudo-wave theory \citep{delannee2008}. In the past, the association of CBFs with type II and type III bursts has been used as evidence against this pseudo-wave interpretation and more in favor of the MHD wave paradigm \citep{warmuth2004b, grechnev2011}. This study reveals that the CBF in this event was indeed closely associated with shock radio activity positioned on the flanks of an expanding CME. This kind of behavior has been suggested before, but never directly imaged \citep{kozarev2011, feng2012, feng2013}. It shows how a combination of radio and EUV imaging can reveal the evolution of plasmoid driven shocks in the solar atmosphere \citep{bain2012}.

Of further interest in this study is the likelihood of a quasi-perpendicular orientation of the shock, as revealed by the PFSS extrapolation. 
Quasi-perpendicularity is an essential aspect of the shock drift acceleration (SDA) mechanism \citep{ball2001}, a process believed to be responsible for particle acceleration in planetary magnetospheres \citep{wu1984} and solar radio bursts \citep{holman1983}.
This mechanism involves an adiabatic reflection of particles from the shock, with the energy gain sourced in the $\mathbf{V} \times \mathbf{B}$ electric field, where $\mathbf{V}$ and $\mathbf{B}$ are the upstream flow speed and magnetic field, respectively. A single reflection from the shock has limited energy gain, however multiple reflections may produce relativistic energies, which is particularly important for low Mach number shocks such as that reported here ($M_A =2.4^{+0.7}_{-0.8}$) and in \citep{guo2012}. This multiple reflection process may be explained by inhomogeneity in the shock front, a characteristic usually known as \textquoteleft rippling' \citep{zlobec1993, vandas2011}. 2D hybrid simulations show that rippling is brought about by an instability \citep{burgess2006} and resembles a standing-wave mode of the shock surface \citep{lowe2003}. The presence of ripples can lead to a quasi-sinusoidal variation in shock-normal orientation with respect to the upstream magnetic field. Since the efficiency of SDA requires quasi-perpendicularity, there will be sites on the shock front that provide efficient acceleration and sites that do not $-$ a structure that may lead to magnetic trapping and multiple reflections \citep{zlobec1993}, hence producing higher energy particle acceleration. 

\begin{figure*}
\includegraphics[scale=0.6, angle=0, trim=3cm 3.5cm 4cm 0cm]{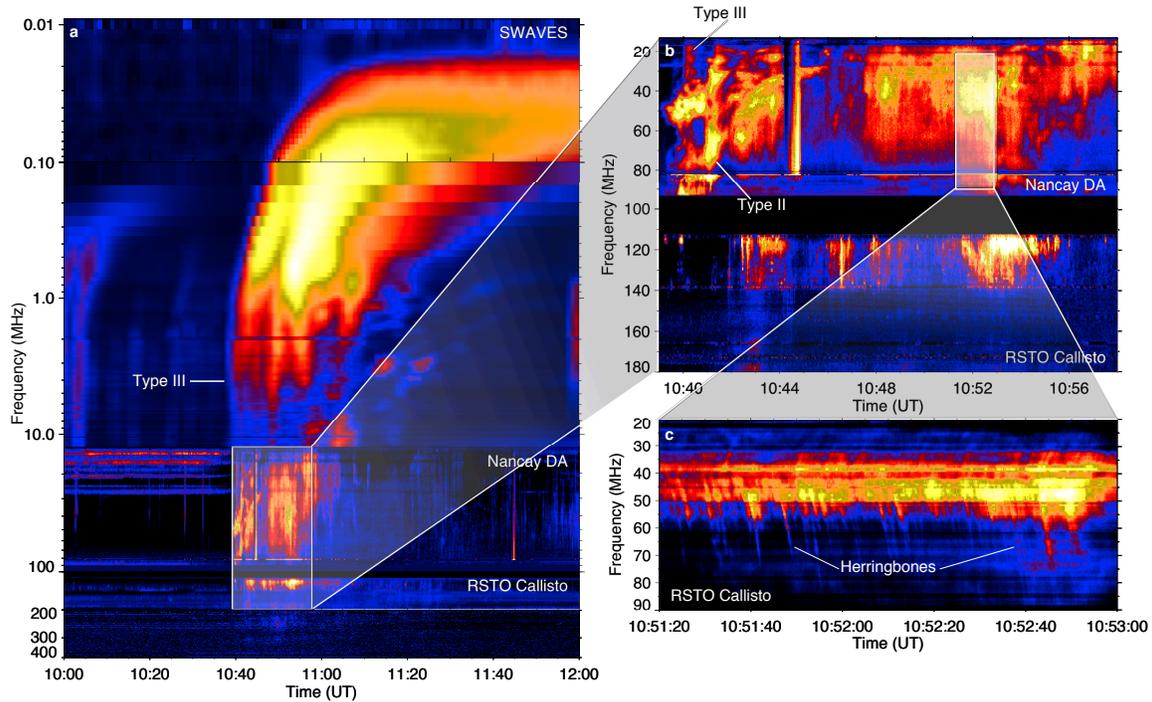}
\caption {Radio dynamic spectra from STEREO-B/WAVES (0.01--16\,MHz), Nan\c{c}ay DA (20--90\,MHz), and RSTO eCallisto (10--400\,MHz). The type II radio burst is indicated in {\bf b}, with both fundamental and harmonic emission observable. This shock signature is characterized by two emission bands drifting slowly ($\sim$-0.2\,MHz\,s$^{-1}$) toward lower frequency over time. Negative frequency drift in dynamic spectra is a result of plasma emission occurring at decreasing density with respect to time. This is due to the emission exciter traveling to larger heights (lower densities) in the solar atmosphere; as the density drops, so too does the frequency of plasma emission.  The type III bursts are indicated in {\bf a},{\bf b}, while herringbones are shown in {\bf c}. Each herringbone or  \textquoteleft spike' is indicative of an electron beam traveling away from the shock. Note that all of the radio activity from {\bf a-c} is indicative of either particle acceleration or a plasma shock in the corona. The start and stop times of this radio activity in these dynamic spectra show good temporal correspondence with the start/stop times of the activity in Fig.~\ref{fig:figure_aia_nrh_c2}. This is especially apparent for the features between 100--200\,MHz; the correspondence between dynamic spectral activity and imaging activity is most apparent in Supplementary Movie 1}
\label{fig:dyn_spec}
\end{figure*}

\begin{figure}[!h]
\includegraphics[scale=0.6, trim=4cm 0cm 2cm 2cm]{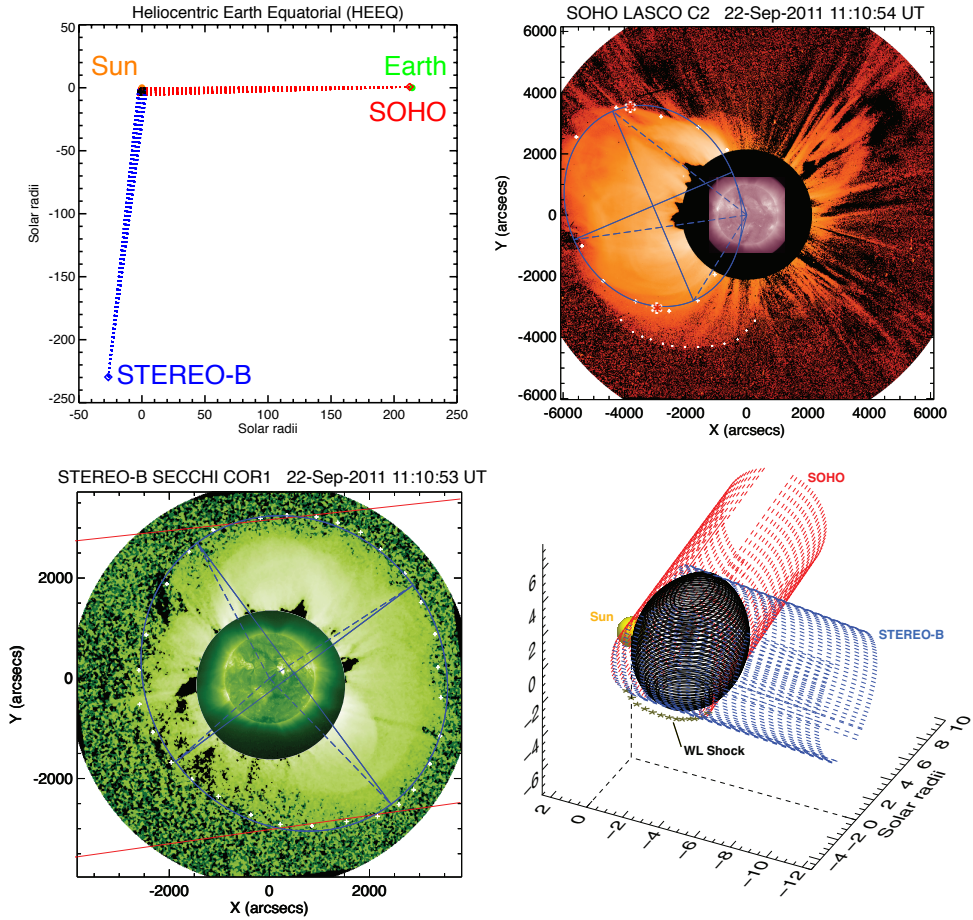}
\caption{White-light CME observations and 3D reconstruction of the CME front. {\bf a} Top-down view of the Heliocentric Earth Equatorial (HEEQ) system, showing the separations and locations of STEREO-tB and SOHO spacecraft with respect to the Sun. {\bf b} LASCO/C2 base-differenced image of the CME (logged intensity scale), with AIA 21.1\,nm image inset. White crosses indicate a point-and-click along the CME front with a corresponding ellipse fit in blue, where the solid lines indicate the major and minor axes, while dashed lines indicate the apex points back toward the Sun centre. The white circles indicate the white-light shock. The red asterisk points indicate the northern and southern flanks of the CME. {\bf c}, Base difference image of the CME from the COR1-B coronagraph, with a corresponding ellipse fit and EUVI 19.5\,nm image inset. The red lines are the red asterisk points in {\bf b} projected as lines-of-sight across the COR1 field of view. {\bf d} 3D reconstruction of the CME with the white light shock indicated on the plane of sky (only 2D information is available for this feature). The red dotted lines are the projected points from the ellipse on the C2 image, and the blue dotted lines are the projected points from the ellipse on the COR1 image. The black ellipses are those inscribed in the resulting quadrilateral slices via the elliptical tie-pointing method for 3D CME reconstruction, as described in \cite{byrne2010}.}
\label{fig:wl_shock}
\end{figure}

The presence of ripples can produce quasi-periodic herringbones in three ways. Firstly, it makes SDA more efficient and capable of producing the observed herringbone energies, especially when particle scattering is considered \citep{burgess2006}. Secondly, the periodic spatial variation in the acceleration efficiency of the shock could explain the bursty and quasi-periodic nature of the herringbones (Supplementary Figure 1). \cite{guo2010} suggest that shock front inhomogeneity brought about by MHD turbulence is a possible explanation of bursty herringbones. \cite{schmidt2012} produced a detailed model of SDA from a rippled shock, specifically on the flanks of an expanding CME. Their results suggest that herringbones could be produced by accelerated electrons at spatially intermittent regions of quasi-perpendicularity on a rippled shock surface. Thirdly, \citep{burgess2006} also predicts that, if rippling is present, the upstream and downstream electron beams should have similar energies, which is not predicted for a uniform or \textquoteleft smooth' shock. A sample of the oppositely drifting herringbones in Fig.~\ref{fig:dyn_spec}c shows that positive and negative frequency drifts are both $\sim$5\,MHz\,s$^{-1}$, revealing that the upstream and downstream populations have similar energies (although we note the possibility that both positive and negative drifting herringbone features may be accelerated upstream, as suggested by the schematic of \citep{zlobec1993}). There is also the possibility that the herringbones may be associated with a termination shock of a reconnection outflow occurring behind the CME \citep{aurass2004}. In such a scenario, this shock would have a more indirect relationship with the CME propagation. However, the imaged radio source shows a good temporal correspondence with the shock activity in the dynamic spectra, especially between 10:52$-$10:56\,UT, suggesting the particle acceleration indicated in the spectra shares a close relationship with the propagating source.

Our observations reveal the need for a more detailed modeling of herringbone solar radio bursts. The quasi-periodic behavior of herringbones provides a possible direct measure of shock inhomogeneity and the spatial scales over which the magnetic field varies in the shock and ambient corona; it may also provide a measure of the turbulence in these plasma flows \citep{guo2010}. In the future, high cadence EUV imaging from SDO, combined with sensitive radio imaging-spectroscopy observations from instruments such as the Low Frequency Array \citep[LOFAR;][]{vanHaarlem2013}, will reveal unprecedented detail of plasma shocks and their role in particle acceleration. This may reveal the fundamental nature of a plasma shock process that is universal, but currently impossible to directly observe in any other area of astrophysics. 
\clearpage

\clearpage


\subsection*{Acknowledgments}
We would like to thank NASA's SDO, STEREO, and ESA/NASA's SOHO teams, and the Nan\c{c}ay Radio Astronomy Observatory for providing open access to their data. Financial support of E.P.C. was provided by the Irish Research Council Embark Initiative. D.M.L. is funded by the European Commission's Seventh Framework Programme under the grant agreement No. 284461 (eHEROES project). J.P.B. is supported by SHINE grant 0962716 and NASA grants NNX08AJ07G and NNX13AG11G to the Institute for Astronomy. P.Z is currently funded under the Trinity College Dublin Innovation Academy Bursary. D.S.B. is funded under the ESA PRODEX programme. We would also like to extend thanks to the Birr Scientific and Heritage Foundation, supported by the Earl of Rosse. Special thanks is extended to Christian Monstein for his support in setting up the Callisto spectrometers and Jasmina Magdaleni\'{c} for very useful scientific discussions.

\subsection*{Author contributions}
E.P.C. performed the data analysis of the radio source kinematics, the radio burst analysis, the Alfv\'{e}n Mach number calculations, and the in-situ particle analysis. E.P.C. also wrote the article. D.M.L. performed the data analysis of the coronal bright front and gave constructive advice on the writing of the article. J.P.B. performed the 3D reconstruction of the CME and gave advice on the white-light shock analysis section. P.Z. provided the density maps, and D.S.B. provided the magnetic field maps that were used in the radio source and CBF Mach number calculations. J.M. installed the electronic systems at RSTO. P.T.G. conceived of the project and guided data analysis and writing of the article.


\def\araa{ARA\&A}%
\def\apj{ApJ}%
\def\apjl{ApJ}%
\def\aap{A\&A}%
\def\aaps{A\&AS}%
\def\icarus{Icarus}%
\def\mnras{MNRAS}%
\def\pasa{PASA}%
\def\solphys{Sol.~Phys.}%
\def\ssr{Space~Sci.~Rev.}%
\def\nat{Nature}%
\def\aplett{Astrophys.~Lett.}%
\def\grl{Geophys.~Res.~Lett.}%
\def\jgr{J.~Geophys.~Res.}%


\bibliographystyle{naturemag}

\end{document}